# Visualizing uniform lattice-scale pair density wave in single-layer FeSe/SrTiO$_3$ films

Yao Zhang[1], Lianzhi Yang[1], Chaofei Liu[1], Wenhao Zhang[1,*], Ying-Shuang Fu[1,*]

1. School of Physics and Wuhan National High Magnetic Field Center, Huazhong University of Science and Technology, Wuhan 430074, China

Emails: *wenhaozhang@hust.edu.cn, *yfu@hust.edu.cn

**Typical BCS superconductors are microscopically homogeneous in real space governed by the coherent Cooper pairs with high phase stiffness of superfluid density, which is characterized by a coherence length. However, a periodic oscillation of superconducting order parameter may develop driven by breaking the time-reversal or translational invariance. To date, such modulated orders were specific to each material systems, with a periodicity much larger than the lattice constant. Here we report the direct observation of a uniform lattice-scale pair density wave (PDW) in single-layer FeSe/SrTiO$_3$ films, enforced by peculiar interfacial structure of crystal symmetries breaking. Our spectroscopic imaging scanning tunneling microscopy unravels a spatial modulation of Cooper-pairing gap within a single unit-cell, depending on inequivalent atomic sites. Prominent periodic variation of superfluid density is visualized via Josephson current by a superconducting tip, indicating a real-space oscillation of phase stiffness. Such a lattice-scale superconducting modulation, which coexists with a larger length scale of PDW order, indicates the lattice-scale variation of both pairing strength and phase stiffness. Our findings provide new insights into the intertwined density-wave orders of quasiparticle character in correlated electronic systems, and provoke future studies on the unconventional pairing interaction and phase stiffness in the two-dimensional limit.**

Superconductivity is a macroscopic quantum phenomenon where condensed Cooper pairs exhibit long-range coherence in equilibrium, protected by time-reversal and spatial inversion symmetries. The superconducting order parameter is thus spatially homogeneous over a superconducting coherence length (the size of Cooper pairs), which typically ranges from several to hundreds of nanometers [1-3]. By breaking either of these two symmetries, the condensations of Cooper pairs are predicted to exhibit nontrivial spin configurations in momentum space. This gives rise to non-uniform superconductivity with a periodic modulation of electron density in real space. To be specific, the absence of time-reversal symmetry can lead to a finite momentum of Cooper pairs via Zeeman splitting effect, the so-called FFLO state that has been reported in systems including heavy-fermion materials, Ising superconductor, and $Sr_2RuO_4$ [4-9]. Alternatively, breaking space-group symmetries can also develop a spatially periodic oscillation of coherent Cooper-pair density, known as PDW. The PDW order is an exotic superconducting state associated with strong electronic correlations, favoring spatially oscillating electron-density dictated by the finite-momentum pairing [10-12].

As a phase of matter in terms of broken symmetries, PDW can simultaneously induce other electronic instabilities of charge/spin density wave and nematic phase in strongly correlated systems [13-15]. By sharing the same wavevector, these accompanying orders are entangled with elusive interplay, resulting in a variety of complicated patterns that are difficult to distinguish. Nevertheless, the wavelength of the PDW order, determined by the wavevector Q of non-zero-momentum Cooper pairs, can extend to about 2-8 times the in-plane lattice constant ($a_0$), commensurate or incommensurate with the underlying lattice, which is comparable to the coherence length [16,17]. The zero-field PDW order could be a widespread phenomenon than previously thought in a wide variety of emerging quantum materials, as recently reported in high-Tc cuprates, iron-based superconductors, kagome compounds, transition-metal dichalcogenides, triplet superconductors, as well as in systems down to the two-dimensional (2D) limit [18-23].

A single-layer FeSe film grown on $SrTiO_3$ substrate (FeSe/STO) is an unconventional superconductor in the 2D limit, where the phase fluctuations become more significant with a lower superfluid density ($\rho_s$) of coherent pairs [24-27], as revealed by the Berezinskii-Kosterlitz-Thouless transition [28,29]. Given the significantly enhanced superconducting pairing gap ($\Delta$) and relatively low superfluid density, the single-layer FeSe/STO system is proven to delicately reside in the BCS-

BEC crossover regime [30-33]. Recent STM study also reports the observation of PDW localized at lattice boundaries of Fe(Se,Te)/STO films, a daughter phase of FeSe/STO, due to translational symmetry breaking [23]. As such, FeSe/STO should be a promising platform to investigate the spatial variation of both the superconducting paring strength and phase stiffness, if the superconducting order parameter varies periodically in space. Specifically, if its lattice symmetry is broken in multiple ways within a unit-cell (UC), the non-uniform superconductivity is expected to be more pronounced among different atomic sites. This can contribute to an intra-UC modulation of density-wave order as a lattice-scale PDW, which however is challenging to be probed in experiments due to the negligible fluctuations in superfluid phase stiffness.

In this study, by using single-electron tunneling of the atomic-resolution STM/STS measurements, we report the atomic observation of an intra-UC superconducting modulation in a single-layer FeSe/STO film. Two sets of nonequivalent atoms, including both the Fe and Se sublattices, exhibit a site-dependent oscillation of Cooper-pair with a $\sqrt{2}a_{Fe}$ period, where $a_{Fe}$ is the lattice constant of Fe primitive cell. Further Josephson STM experiments of Cooper-pair tunnelling validate the nonequivalence of sublattice by spatially imaging a real-space variation of superfluid density. Both intra-UC and inter-UC PDW with different wavelengths can coexist, which depends on different combinations of multiple symmetries breaking in FeSe crystal lattices.

**Symmetries breaking by the interfacial effect**

High-quality single-layer FeSe films were epitaxially grown onto a clean Nb-STO(001) substrate as our previous studies (see Methods) [28]. Figure 1(c) shows a typical STM topography of the FeSe/STO surface with a Se-1×1 termination in square lattice. Simultaneously, a unidirectional 2×1 reconstruction is commonly observed with weak but visible striped patterns, which are more clearly seen from the fast Fourier transform (FFT) spots as inserted. Similar surface reconstruction has been reported previously, which is originated from the ordered O vacancies of STO substrate introduced by the over-rich Se flux during growth [34]. Figures 1(a),(b) illustrate the structural model of single FeSe layer on the O-deficient substrate, which is characterized by alternately missing O atoms along a-axis. During the growth, the foreign Se atoms preferentially substitute the sites of O vacancies to form the bottom Se layer of triple-layer FeSe. As a result, both the interfacial Ti-Se and Se-Fe-Se bonds in FeSe are distorted with distinctive bonding lengths and angles, generally generating the inversion symmetry breaking. The two Fe atoms in Fe sublattice, which originally reside in the same atomic layer, then become nonequivalent with different bond environments.

Due to the convolved lattice structure and electronic states, our STM can not only resolve the topmost Se atoms, but also detect the sublayer Fe primitive cell, which has not been achieved previously. Figs. 1(c),(d) compare the exact occupations of both Se and Fe sites in the same field of view, but are acquired under different bias voltages. There exhibits a relative 45° rotation between the directions of the Se and Fe sublattices, whereas endowing similar FFT architecture with different intensities. Remarkably, the nonequivalent Fe atoms within a UC, induced by the interfacial effect, can be locally resolved with distinct bonding features. As seen from the close-up image of inset in Fig. 1(d), we can resolve dissimilar contrasts of nonequivalent Fe atoms, which are imaged by stronger bonding feature linked to the topmost Se sites, unambiguously verifying the broken $C_4$ rotational symmetry. Combined with the inversion symmetry breaking, both the Fe and Se atom within a UC can be sorted into two classes: Se-1 vs. Se-2, and Fe-1 vs. Fe-2, respectively.

**Atomic observation of superconducting gap modulation**

The nonequivalence of atoms is examined by the typical superconducting spectra on each site compared in Fig. 1(e), which are collected at 1.6 K, far below the superconducting transition temperature above 60 K [35]. All STS spectra are U-shaped and exhibit a double-gap structure with pairs of coherence peaks at $\Delta_1 = \pm 15$ and $\Delta_2 = \pm 9$ mV, respectively, well comparable with previous values [24,29,36]. The tunnelling conductance in the energy range of ±5 mV is completely zero, suggesting its fully gapped nature with high sample quality. Upon slightly increasing to larger energies, there emerge quasiparticle excitations as a pair of shoulders or kinks at ±6.5 mV in the superconducting spectra, which always obey the particle-hole symmetry in analogy to $\Delta_1$ and $\Delta_2$. We thus denote them as $\Delta^*$. The multiple-gap structure in FeSe/STO remains elusive, and is usually discussed to be a result of the complex interplay including the multiple Fe-3d orbital nature, as well as their inter-orbital interaction, band anisotropy, electron-phonon coupling, and the interfacial effect [37-40]. The superconducting discrepancies between the same kind of atoms are more amplified by showing their spectroscopic difference for each of the two sublattices, red for Se and blue for Fe sites, respectively, in the bottom panel of Fig. 1(e). Apparently, the most prominent variations occur in the energy range of ±[5,20] mV concerning the coherent quasiparticles. We thus observe very small variations of superconductivity among different atomic sites inside a single UC, especially near the energies around $\Delta_1$, $\Delta_2$, and $\Delta^*$.

To gain further insight into the variations of superconducting spectra, we scrutinize their high-resolution STS linecuts from Fig. 2(a) measured along the y-axis, which only cross either the inequivalent Se-Se or Fe-Fe atoms. As shown in Figs. 2(b)-(f), an oscillatory feature is well observed along both trajectories, showing nonequivalence between the Se-1 (Fe-1) and Se-2 (Fe-2) sites. Such an electronic modulation is robust against various tip-sample distances (Fig. S1), thus excluding any possible tip effect. The spectroscopic differences between Se and Fe sites are also presented in Fig. S2, with linecuts along the nearest neighboring Fe-Se direction (a/b axis). To quantitatively analyze the superconducting modulation featured by the coherence peaks, we extract the gap size ($\Delta_1$) and its coherence peak height (CPH) in a spatial evolution plot in Figs. 2(d) and (f). Clearly, the gap size exhibits spatial modulation in a period of $2a_{Fe}$ for both Se and Fe sublattices, as denoted in Fig. 2(a). Similarly, the intensities of CPH also illustrate the same modulation periodicity. The site-dependent variations are irrelevant to the spectral background (Fig. S3), indicating a robust modulation of the superconducting order parameter within the UC. Interestingly, the location-dependent $\Delta$ and the associated CPH forms an anti-phase relationship for both trajectories, as indicated by the vertical dashed lines. Moreover, $\Delta$ displays local maxima at the Se-2 and Fe-2 sites, respectively, and local minimal at the relatively farther Se-1 and Fe-1 sites. Note that the variation of superconducting gap size is ~1 meV, and are weakened for the Fe linecut compared with the Se case. This may be attributed to the more notable asymmetry in structure between the top and bottom layers of Se, whereas there is less nonequivalence among the otherwise coplanar Fe sublattice.

**Visualization of the superfluid density modulation**

It has been reported that the CPH can be regarded as a local indicator of the quasiparticle strength of superconductivity, which is conjectured to phenomenologically evaluate the level of superfluid density [41]. As such, the CPH of quasiparticle character can reflect the degree of coherence for condensed Cooper pairs, known as the phase stiffness in unconventional superconductors [42]. More directly, the superfluid density ($\rho_s$) can be accessible via measuring Josephson current assisted by inelastic Cooper-pair tunnelling near the Fermi level. We then conduct spatial imaging of the superfluid density by employing Josephson STM techniques with a superconducting Nb tip. Fig. 2(g) is the typical Josephson tunnelling spectra of current-voltage and differential conductance characteristics for different tip-sample distances. Clearly, there emerges a progressive local maximum current ($I_{max}$ in I-V) and a zero-bias peak (ZBP) in dI/dV for decreasing the normal-state resistance ($R_N$). These are the typical hallmarks for the expected Josephson current of Cooper-pair tunnelling [43,44]. In principle, the conductance at Fermi energy (ZBP) is proportional to the value of $I_{max}$, which is also proportional to the square of the critical supercurrent $I_C$, that is, ZBP $\propto I_{max} \propto (I_C)^2$ (Fig. S4). The superfluid density for sample can be determined through the relationship: $\rho_s \propto (I_C \times R_N)^2$, where the electron density of superconducting tip keeps constant. In this regard, the quantities of both ZBP and $I_{max}$ offer experimental means to measure the local superfluid density in real space [21].

Fig. 2(h) is the spatial distribution of differential Josephson spectra as a function of position taken along the Fe-Fe directions, similar as Fig. 2(e). Given that $R_N$ always remains spatially constant for each line spectra, the superfluid density can be simply measured from the ZBP height or $I_C$ from the I-V curves, whose spatial variations are extracted in Fig. 2(i). Herein pronounced contrast between the inequivalent of Fe-1 and Fe-2 sites are more clearly resolved, nicely coinciding with the oscillatory behavior of the gap size and CPH commensurate with the lattice. The spatial distributions for $\rho_s$ along the Se-Se direction can be seen in Fig. S5, providing strong evidences for the intra-UC modulations of both the Cooper-pair density and superfluid density. Considering the structurally inequivalence of Se and Fe sublattices, our observations of stie-dependent modulation within a UC are likely induced by the combination of breaking both the $C_4$ rotational and inversion symmetries in the single-layer FeSe/STO films.

Such spatial variations of the Cooper-pair and superfluid density can be more clearly visualized from the spectroscopic imaging STM measurements with atomic resolution. Figs. 3(a)-(c) are the topographic image together with its associated two conductance mappings at energies of $\Delta_1$ and $\Delta_2$. The Se (Fe) sublattice with enhanced conductance is achieved at the energy of $\Delta_1$ ($\Delta_2$), which is attributed to the $d_{xy}$ ($d_{xz}/d_{yz}$) orbital characters of Fe atoms, respectively [38]. Considering the p-orbital electrons of the Se atoms, the reversed maps between $\Delta_1$ and $\Delta_2$ is well consistent with the Se-p and Fe-d orbital hybridization: Se sites are contributed by the inter-orbital interaction of p-$d_{xy}$ orbitals at $\Delta_1$, whereas there is little overlap between p and $d_{xz}/d_{yz}$ orbitals, thus visualizing the Fe sites at $\Delta_2$. We thus speculate that the p-d hybridization should play an essential role of Cooper pairing interactions in our observed superconducting modulation.

We further extract the gap ($\Delta_1$) and CPH at each pixel in the identical field of view, as displayed in Figs. 3(d),(e), where the spatial modulations are more visible from the Fourier filtered maps (Fig. S6). Although the overall gap size distribution is much wider (~5 meV from the histogram inserted) than that from the line spectra in Fig. 2 (~1 meV) due to inhomogeneities, both the gap amplitude and quasiparticle intensities exhibit a remarkable correlation with a common period of $\sqrt{2}a_{Fe}$. Quantitatively, the correlation coefficient between the maps of gap size and CPH is calculated to be -0.30 (Fig. 3(f)), indicative of a distinct evolution for the amplitude of energy gap and the phase stiffness of Cooper-pair breaking.

**Coexistence of inter-UC PDW order and intra-UC modulation**

Since the 2×1 interfacial structure breaks the $C_4$ rotational symmetry into $C_2$, it inevitably introduces different domain structures along mutually perpendicular directions. As displayed in Fig. 4(a), the optional orientation of 2×1 stripes naturally create a domain wall region with continuous Se atoms, but with a 1/2-unit-cell lattice shift across the boundaries (Fig. S7). Previous study has uncovered a PDW state with a period of $3.6a_{Fe}$ at the domain walls of Fe(Te,Se)/STO, which is explained by the strong Rashba and Dresselhaus spin-orbit coupling due to the broken spatial symmetries [23]. To differentiate the two types of PDW, we define such large period PDW as inter-UC type, and our lattice-scale PDW of $\sqrt{2}a_{Fe}$ period as intra-UC type. As presented in Fig. 4(c), we also observe a similar inter-UC PDW, but with a different period of $5.4a_{Fe}$ from the conductance mapping along the Fe-Fe direction. Meanwhile, there also emerge the intra-UC PDW modulation of gap size and CPH with similar features as discussed in Figs. 2 and 3. The coexistence of inter-UC and intra-UC PDW modulations can be well resolved from a detailed STS line spectra along the domain wall in Fig. 4(d). While the inter-UC PDW order is only observed in a very narrow energy windows near $\Delta^*$ (±[4.5, 7.8] mV), the intra-UC PDW can exist for wider energies that quasiparticles survive below the pair-breaking gap. As compared from the conductance profiles near $\Delta^*$ and $\Delta_1$ in Fig. 4(e), while the former is a superimposition of $\sqrt{2}a_{Fe}$ and $5.4a_{Fe}$ PDW, only the $\sqrt{2}a_{Fe}$ modulation is preserved in the latter case, which is verified from the intensities of the FFT amplitudes in Fig. 4(f).

To decouple these two electronic orders, we perform the STS mappings at the energy range for the emergence of intra-UC PDW order in the interior of individual domains, where the inter-UC modulation is absent off the domain wall. As compared in Fig. 4(g), the LDOS manifests a striped pattern with conductance maximum located at the Fe atoms, accompanied by its orientation perpendicular to the direction of the 2×1 reconstruction (Fig. S8). Moreover, there emerges a π-phase shift relationship of stripes between the occupied and unoccupied states, namely, enhanced conductance at the Fe-2 (Fe-1) site for +(-) 5.4 mV. The LDOS imbalance strongly suggests the nonequivalence in electronic properties between Fe-1 and Fe-2 sites. This may be understood from the rotational symmetry breaking by the interface, particularly the 2×1 reconstruction, which causes imbalanced bonding environments for the inequivalent Fe atoms, as discussed in Fig. S9. The asymmetric pairing strengths, mediated by the nearest neighboring exchange interactions, further lift the sublattice degeneracy with varied hybridization between the Se-p and Fe-d orbitals. In this light, the structural symmetry breaking results in the site-dependent modulation of the superconducting pairing interaction within a lattice scale.

**Discussion**

Our observations of intra-UC modulation are entirely different from the inter-UC PDW order driven by the breaking of translational symmetry inside particular domain walls in Fe(Te,Se)/STO films [23]. Compared to previous Fe(Te,Se) study, FeSe/STO not only hosts the same advantages of high-$T_C$ superconductivity and nontrivial topological properties, but is also superior in three ways: (i) FeSe is structurally more simple without atomic intermixing, where the randomly distributed Se/Te atoms in both the top and bottom chalcogen layers of Fe(Te,Se) makes the symmetry-breaking much complicated. (ii) The stoichiometric variations in Fe(Te,Se) will inevitably introduce local inhomogeneity in electronic structure, which severely hinders resolving other possible spatial oscillations at atomic scale. (iii) The absence of hole pocket in the Brillouin zone center of Fermi surface, as reported by previous phase-referenced quasiparticle interference and angle resolved photoemission spectroscopy measurements, directly quenches the nesting vector between the Γ and M Fermi pockets thus suppresses nematic phase [26]. All these privileges make FeSe/STO more applicable to unravel the intra-UC superconducting modulations. Here in our case, while the PDW order with larger period still survives at the domain boundaries, the elaborate $\sqrt{2}a_{Fe}$-modulation is driven by further breaking of rotational and inversion symmetries in the specific interfacial structure.

Finally, we discuss the possible scenario for the $\sqrt{2}a_{Fe}$ PDW order. It has been argued that the CDW order in NbSe$_2$ will generate a higher-order periodic potential promoting the electron pairing with non-zero center-of-mass momentum, whose wavevectors are the same as the reciprocal lattices of the CDW state [21]. Meanwhile, the leading-order quasiparticle density of states is expected to follow the periodicity of crystal lattice according to the Bloch's theorem, which can also apply to superconductivity. In the presence of multiple symmetries breaking in single-layer FeSe/STO, the peculiar interfacial structure will induce strong spin-orbit coupling and facilitate the inter-orbital hybridization of Se-p and Fe-d orbitals with anisotropic bond environment, thus locally altering the paring strength with varied superfluid density. Similarly, the asymmetric pairing strength further yields a periodic potential from the crystal lattice, modulating the electron tunneling of quasiparticle character and pair-density. This is propitious to stabilize a finite momentum electron-pairing with the Bragg wavevectors (schematically shown in Fig. S10), leading to our observations for real-space oscillation of the superconducting order parameter commensurate with the lattice periodicity.

In summary, we directly visualize a spatial modulation of the superconducting order parameter at the atomic scale, whose wavelength coincides with the lattice period. The intra-UC oscillation is evidenced by both single-particle and Cooper-pairing tunneling measurements, simultaneously demonstrating the spatial distributions of gap magnitude and superfluid density with coherent quasiparticles. This $\sqrt{2}a_{Fe}$-modulation may involve the bond-sensitive inter-orbital hybridization with local paring strength, which relies on different space symmetries breaking and is distinguished from the coexisting inter-UC PDW ordering. Future investigations on the role of the exotic topological excitations, as well as the responses of both nonmagnetic (Se vacancy) and magnetic impurities (Fe vacancy) coupled to these modulations, may help to elucidate the unconventional pairing interaction in such a strongly correlated superconductor.

Note: During the preparation and submission process of our manuscript, we are aware of two parallel works of observing a similar intra-UC PDW state, which survives on a more complex Fe(Se,Te) system [45,46]. However, there presents many differences regarding the broken symmetries and atomic inequivalences, which may be attributed to the variations in specific materials and experimental circumstances.

**Acknowledgements**

This work is funded by the National Key Research and Development Program of China (Grants No. 2022YFA1402400, 2018YFA0307000), the National Science Foundation of China (Grants No. 92265201, U20A6002, 12174131, 11774105, 11874161), the Natural Science Foundation of Hubei (2022CFB033) and Knowledge Innovation Program of Wuhan-Basic Research (2023010201010056).

**References**

[1] Dagotto, E. Correlated electrons in high-temperature superconductors. *Rev. Mod. Phys.* **66**, 763–840 (1994).

[2] Kallin, C. & Berlinsky, J. Chiral superconductors. *Rep. Prog. Phys.* **79**, 054502 (2016).

[3] Stewart, G. R. Superconductivity in iron compounds. *Rev. Mod. Phys.* **83**, 1589–1652 (2011).

[4] Kenzelmann, M. Exotic magnetic states in Pauli-limited superconductors. *Rep. Prog. Phys.* **80**, 034501 (2017).

[5] Kumagai, K. *et al.* Fulde-Ferrell-Larkin-Ovchinnikov State in a Perpendicular Field of Quasi-Two-Dimensional $CeCoIn_5$. *Phys. Rev. Lett.* **97**, 227002 (2006).

[6] Wan, P. *et al.* Orbital Fulde–Ferrell–Larkin–Ovchinnikov state in an Ising superconductor. *Nature* **619**, 46–51 (2023).

[7] Kinjo, K. *et al.* Superconducting spin smecticity evidencing the Fulde-Ferrell-Larkin-Ovchinnikov state in $Sr_2RuO_4$. *Science* **376**, 397–400 (2022).

[8] Beyer, R., Bergk, B., Yasin, S., Schlueter, J. A. & Wosnitza, J. Angle-Dependent Evolution of the Fulde-Ferrell-Larkin-Ovchinnikov State in an Organic Superconductor. *Phys. Rev. Lett.* **109**, 027003 (2012).

[9] Cho, C.-woo. *et al.* Thermodynamic evidence for the Fulde–Ferrell–Larkin–Ovchinnikov state in the $KFe_2As_2$ superconductor. *Phys. Rev. Lett.* **119,** 217002 (2017).

[10] Chen, H.-D., Vafek, O., Yazdani, A. & Zhang, S.-C. Pair Density Wave in the Pseudogap State of High Temperature Superconductors. *Phys. Rev. Lett.* **93**, 187002 (2004).

[11] Agterberg, D. F. & Tsunetsugu, H. Dislocations and vortices in pair-density-wave superconductors. *Nat. Phys.* **4**, 639–642 (2008).

[12] Berg, E., Fradkin, E. & Kivelson, S. A. Charge-4e superconductivity from pair-density-wave order in certain high-temperature superconductors. *Nat. Phys.* **5**, 830–833 (2009).

[13] Wang, Y., Agterberg, D. F. & Chubukov, A. Coexistence of Charge-Density-Wave and Pair-Density-Wave Orders in Underdoped Cuprates. *Phys. Rev. Lett.* **114**, 197001 (2015).

[14] Fradkin, E., Kivelson, S. A. & Tranquada, J. M. *Colloquium* : Theory of intertwined orders in high temperature superconductors. *Rev. Mod. Phys.* **87**, 457–482 (2015).

[15] Agterberg, D. F. *et al.* The Physics of Pair-Density Waves: Cuprate Superconductors and Beyond. *Annual Review of Condensed Matter Physics* **11**, 231–270 (2020).

[16] Ruan, W. *et al.* Visualization of the periodic modulation of Cooper pairing in a cuprate superconductor. *Nat. Phys.* **14**, 1178–1182 (2018).

[17] Du, Z. *et al.* Imaging the energy gap modulations of the cuprate pair-density-wave state. *Nature* **580**, 65–70 (2020).

[18] Hamidian, M. H. *et al.* Detection of a Cooper-pair density wave in $Bi_2Sr_2CaCu_2O_{8+x}$. *Nature* **532**, 343–347 (2016).

[19] Zhao, H. *et al.* Smectic pair-density-wave order in $EuRbFe_4As_4$. *Nature* **618**, 940–945 (2023).

[20] Chen, H. *et al.* Roton pair density wave in a strong-coupling kagome superconductor. *Nature* **599**, 222–228 (2021).

[21] Liu, X., Chong, Y. X., Sharma, R. & Davis, J. C. S. Discovery of a Cooper-pair density wave state in a transition-metal dichalcogenide. *Science* **372**, 1447–1452 (2021).

[22] Gu, Q. *et al.* Detection of a pair density wave state in $UTe_2$. *Nature* **618**, 921–927 (2023).

[23] Liu, Y. *et al.* Pair density wave state in a monolayer high-Tc iron-based superconductor. *Nature* **618**, 934–939 (2023).

[24] Wang, Q.-Y. *et al.* Interface-Induced High-Temperature Superconductivity in Single Unit-Cell FeSe Films on SrTiO$_3$. *Chin. Phys. Lett.* **29**, 037402 (2012).

[25] Biswas, P. K. *et al.* Direct evidence of superconductivity and determination of the superfluid density in buried ultrathin FeSe grown on SrTiO$_3$. *Phys. Rev. B* **97**, 174509 (2018).

[26] Huang, D. & Hoffman, J. E. Monolayer FeSe on SrTiO$_3$. *Annual Review of Condensed Matter Physics* **8**, 311–336 (2017).

[27] Yao, G. *et al.* Diamagnetic Response of Potassium-Adsorbed Multilayer FeSe Film. *Phys. Rev. Lett.* **123**, 257001 (2019).

[28] Zhang, W.-H. *et al.* Direct Observation of High-Temperature Superconductivity in One-Unit-Cell FeSe Films. *Chin. Phys. Lett.* **31**, 017401 (2014).

[29] Zhao, D. *et al.* Electronic inhomogeneity and phase fluctuation in one-unit-cell FeSe films. *Nat. Commun.* **15**, 3369 (2024).

[30] Kasahara, S. *et al.* Giant superconducting fluctuations in the compensated semimetal FeSe at the BCS–BEC crossover. *Nat. Commun.* **7**, 12843 (2016).

[31] Lin, H. *et al.* Real-space BCS-BEC crossover in FeSe monolayers. *Phys. Rev. B* **107**, 104517 (2023).

[32] Kasahara, S. *et al.* Evidence for an Fulde-Ferrell-Larkin-Ovchinnikov State with Segmented Vortices in the BCS-BEC-Crossover Superconductor FeSe. *Phys. Rev. Lett.* **124**, 107001 (2020).

[33] Chen, Q., Wang, Z., Boyack, R., Yang, S. & Levin, K. When superconductivity crosses over: From BCS to BEC. *Rev. Mod. Phys.* **96**, 025002 (2024).

[34] Bang, J. *et al.* Atomic and electronic structures of single-layer FeSe on SrTiO$_3$ (001): The role of oxygen deficiency. *Phys. Rev. B* **87**, 220503 (2013).

[35] Zhang, W. *et al.* Interface charge doping effects on superconductivity of single-unit-cell FeSe films on SrTiO$_3$ substrates. *Phys. Rev. B* **89**, 060506 (2014).

[36] Zhang, W. H. *et al.* Effects of Surface Electron Doping and Substrate on the Superconductivity of Epitaxial FeSe Films. *Nano Lett.* **16**, 1969–1973 (2016).

[37] Lee, J. J. *et al.* Interfacial mode coupling as the origin of the enhancement of Tc in FeSe films on SrTiO$_3$. *Nature* **515**, 245–248 (2014).

[38] Zhang, Y. *et al.* Superconducting Gap Anisotropy in Monolayer FeSe Thin Film. *Phys. Rev. Lett.* **117**, 117001 (2016).

[39] Yi, M., Liu, ZK., Zhang, Y. *et al.* Observation of universal strong orbital-dependent correlation effects in iron chalcogenides. *Nat. Commun.* **6**, 7777 (2015).

[40] Huang, D. *et al.* Revealing the Empty-State Electronic Structure of Single-Unit-Cell FeSe/SrTiO$_3$. *Phys. Rev. Lett.* **115**, 017002 (2015).

[41] Cho, D., Bastiaans, K. M., Chatzopoulos, D., Gu, G. D. & Allan, M. P. A strongly inhomogeneous superfluid in an iron-based superconductor. *Nature* **571**, 541–545 (2019).

[42] Emery, V. J. & Kivelson, S. A. Importance of phase fluctuations in superconductors with small superfluid density. *Nature* **374**, 434–437 (1995).

[43] Randeria, M. T., Feldman, B. E., Drozdov, I. K. & Yazdani, A. Scanning Josephson spectroscopy on the atomic scale. *Phys Rev. B* **93**, 161115 (2016).

[44] Graham, M. & Morr, D. K. Imaging the spatial form of a superconducting order parameter via Josephson scanning tunneling spectroscopy. *Phys. Rev. B* **96**, 184501 (2017).

[45] L. Kong, *et al.* Observation of Cooper-pair density modulation state. *arXiv*:2404.10046.

[46] T. Wei, *et al.* Observation of intra-unit-cell superconductivity modulation. *arXiv*:2404.16683.

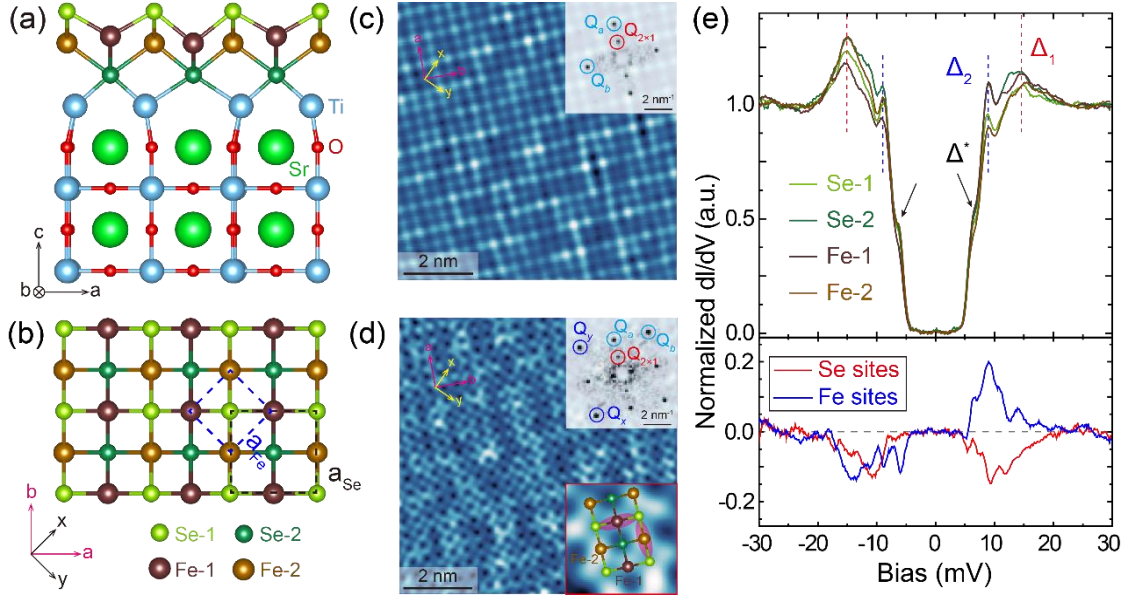

**Figure 1. Crystal and electronic structure of single-layer FeSe/STO films.** (a),(b) The side (a) and top (b) view for the structural schematic of a FeSe layer sitting on the O-deficient STO(001) substrate, respectively, which is terminated by a $TiO_2$ layer with alternately missing O-atom rows along the a-axis. Due to this interfacial structure, both the Fe and Se atoms in a UC are sorted into two classes: the top (bottom) Se layer is donated as Se-1 (Se-2) by light (dark) green balls. The Fe atoms, which are originally coplanar, now become buckled with inequivalent Fe-1 (Fe-2) sites slightly up (down) moving along c-axis, which are presented by dark (light) brown balls. The sublattices for Se and Fe are marked by the dashed black and blue square, with the base vector directions along the a/b and x/y axis respectively. (c),(d) Atomically resolved STM image of the Se (c, $V_b$ = -100 mV, $I_t$ = 100 pA) and Fe (d, $V_b$ = 8 mV, $I_t$ = 100 pA) sublattice, respectively, in the same field of view. Right up insets are the corresponding FFT results, showing both signals for the atomic Bragg and 2×1 reconstruction. Right bottom inset in (d) is a close-up image of Fe sublattice overlaid with an atomic model, guiding to distinguish the inequivalent Fe-1 and Fe-2 sites with bonding different environments (red ellipses). (e) Top panel: typical tunnelling spectra measured exactly on four types of atomic sites within a UC, that is, Se-1, Se-2, Fe-1 and Fe-2 atoms. The superconducting gap sizes, accompanied by the shoulders above the full gap are labelled as $\Delta_1$, $\Delta_2$, and $\Delta^*$, respectively. The very small variations among different STS spectra within the UC always exist after a normalization and background subtraction process, as compared in Fig. S3. Bottom panel: The spectra difference between two sublattices: red for Se-1 ad Se-2, blue for Fe-1 ad Fe-2, respectively. Spectroscopic condition: $V_b$ = 30 mV, $I_t$ = 2 nA, and $V_{mod}$ = 0.3 mV.

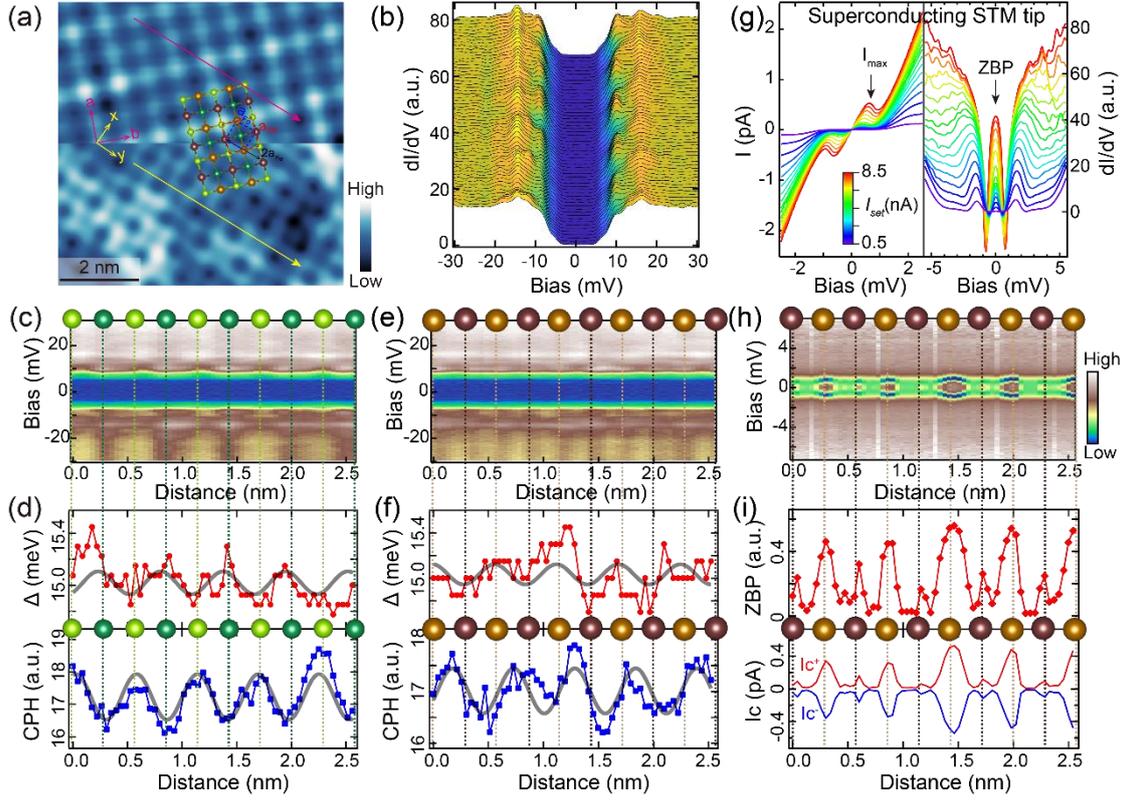

**Figure 2. Intra-UC modulation of superconductivity with a period of lattice constant in single-layer FeSe/STO films.** (a) High resolution STM topography for the FeSe surface imaged with Se (Fe) sublattice in the top (bottom) panel, which is overlaid by a toy model of FeSe crystal structure. (b) Waterfall plot of an STS linecut acquired along the red line of y-axis in (a) across the Se sublattice, which are offset for clarity. $V_b$ = 30 mV, $I_t$ = 1.5 nA, and $V_{mod}$ = 0.3 mV. (c),(d) 2D color plot of (b) in real space, associated with the extracted superconducting gap size ($\Delta_1$) and CPH exhibiting spatial modulation. Gray lines are fitting curves by a cosine function. (e),(f) Similar 2D plot of linecut spectra along the yellow line in (a), only crossing the Fe sublattice. (g) Current-voltage (left) and differential conductance (right) spectra, respectively, measured by a Josephson STM setup for different normal-state resistances. All are acquired at = 6 mV and $V_{mod}$ = 0.05 mV. (h),(i) 2D color plot of the Josephson dI/dV curves across the Fe sublattice, associated with the extracted ZBP and $I_C$ values as a function of positions, exhibiting the same spatial modulation.

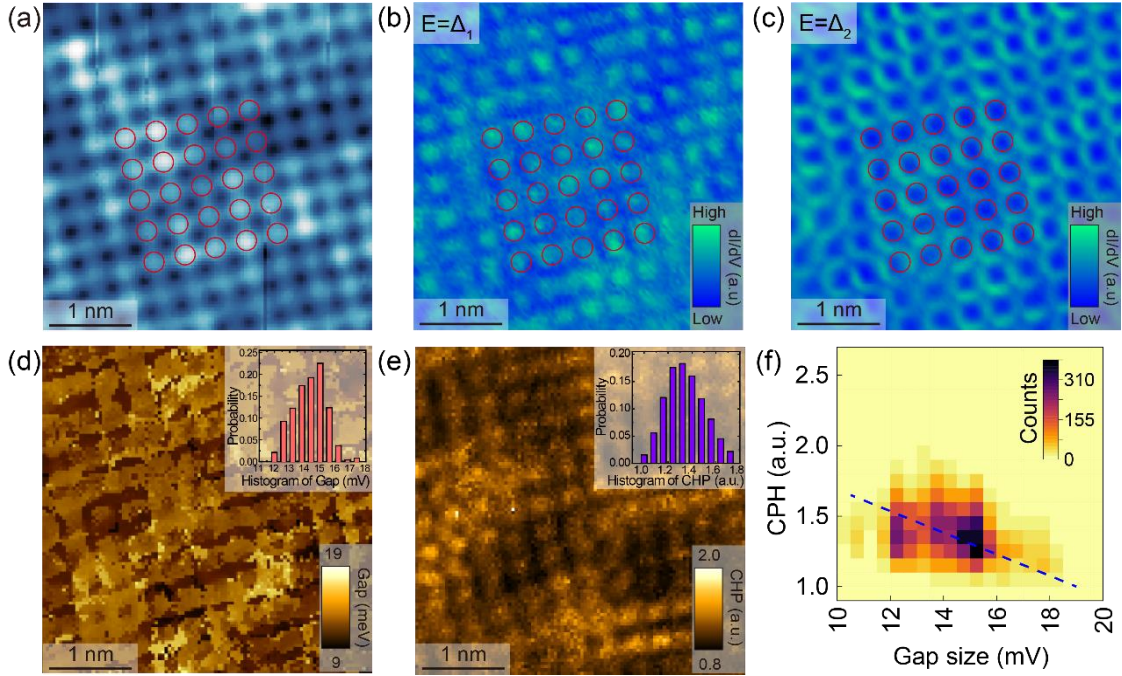

**Figure 3. Visualizing the Cooper-pair and superfluid density modulation in single-layer FeSe/STO films.** (a) STM topographic image for the atomic structure of FeSe surface terminate by Se lattice. (b),(c) STS mappings taken at the same area as in (a), which are acquired at energies of $\Delta_1$ and $\Delta_2$, respectively. The grid of red circles marks the locations of top Se (Se-1) sites. (d),(e) Real space maps of superconducting gap size (b) and CHP (c), respectively, acquired in the identical field of view as (a). Both show the same intra-UC superconducting modulations with a $a_0$ period. Insets are the histograms of gap size and CPH extracted from the spatial distributions. $V_b$ = 30 mV, $I_t$ = 2 nA, and $V_{mod}$ = 0.3 mV with a grid of 625 point/nm$^2$. (f) Correlation between the gap size and CHP, yielding a correlation coefficient of -0.30 (blue dashed line).

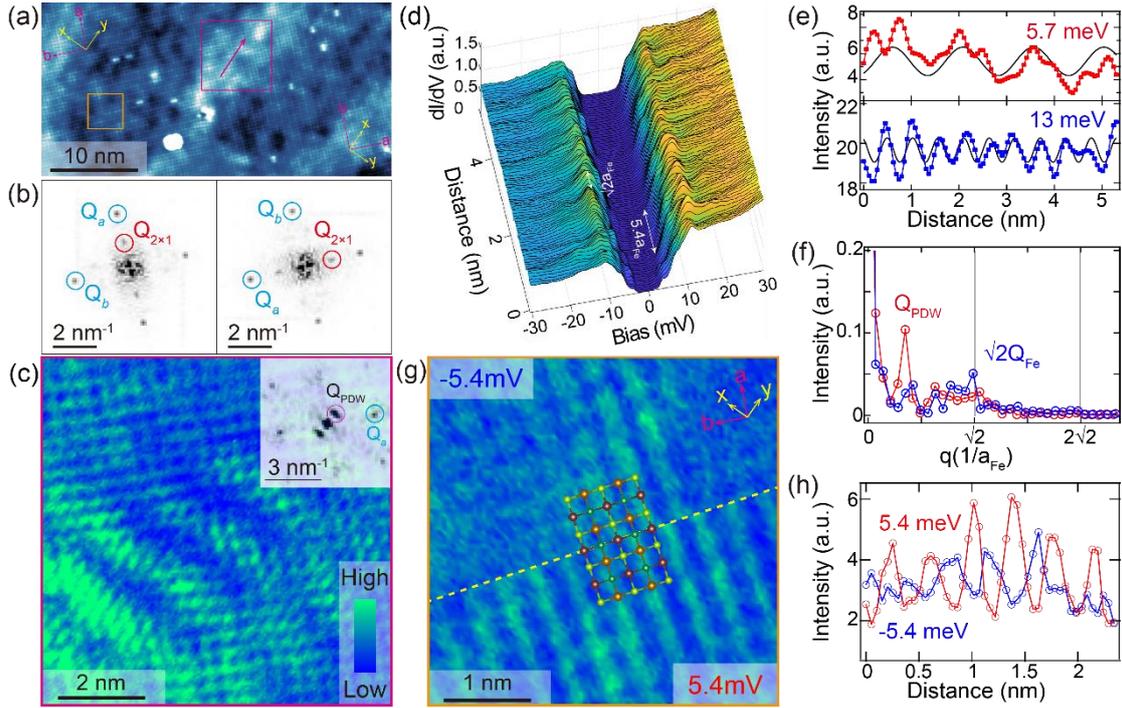

**Figure 4. The coexistence of PDW order and intra-UC modulation in single-layer FeSe/STO films.** (a) STM topographic image for the FeSe surface containing two 2×1 domains that are orientated perpendicularly. The domain wall can be well seen as areas of bright ridges. $V_b$ = -100 mV, $I_t$ = 50p A. (b) FFT images for the left and right 2×1 domains, respectively, showing a relative 90°-orientation to each other. (c) Spatial mapping of differential conductance at -5.7 mV taken from the magenta square on the domain wall of (a), showing a 5.4$a_{Fe}$ PDW modulation along the domain wall of x-axis. $I_t$ = 2 nA, $V_{mod}$ = 0.3 mV with a grid of 350 point/nm$^2$. (d) 3D color plot of the STS spectra taken along the along the magenta line of the domain wall in (a). $V_b$ = 30 mV, $I_t$ = 2 nA, and $V_{mod}$ = 0.3 mV. (e) Conductance profiles extracted along the magenta line at different energies of 5.7 (red) and 13 (blue) mV, respectively. (f) Magnitude of FFT for (e), showing the coexisting $\sqrt{2}a_{Fe}$ and 5.4$a_{Fe}$ PDW modulations. (g) Differential conductance mapping acquired at ±5.4 mV, respectively, with the region selected from the orange square of left domain in (a). The atomically structural model is overlaid on both maps. The striped patterns of the Fe sublattice directly display an inversed contrast between opposite bias voltages, which both orient perpendicularly to the direction of 2×1 reconstruction. $I_t$ = 1.5 nA, and $V_{mod}$ = 0.3 mV with a grid of 400 point/nm$^2$. (h) Conductance profiles extracted from ±5.4 mV at the same locations, which are perpendicular to the stripes in (g), showing an out-of-phase relationship.